\documentclass[preprint,showpacs,preprintnumbers,amsmath,amssymb]{revtex4}
\usepackage{amsmath}
\usepackage{graphicx}
\usepackage{amssymb}
\usepackage{bm}

\begin{document}
\title{Mechanisms producing fissionlike binary fragments in heavy collisions}
\author{A. K. Nasirov}
\altaffiliation{Institute of Nuclear Physics, 100214, Tashkent, Uzbekistan}
\email{nasirov@jinr.ru}
\affiliation{Joint Institute for Nuclear Research, 141980 Dubna, Russia }
\author{G. Giardina}
\affiliation{Dipartimento di Fisica dell' Universit\`a di Messina, 98166 Messina,  and Istituto Nazionale di Fisica Nucleare, Sezione di Catania,  Italy}
\author{G. Mandaglio}
\affiliation{Dipartimento di Fisica dell' Universit\`a di Messina, 98166 Messina,  and Istituto Nazionale di Fisica Nucleare, Sezione di Catania,  Italy}
\author{M. Manganaro}
\affiliation{Dipartimento di Fisica dell' Universit\`a di Messina, 98166 Messina,  and Istituto Nazionale di Fisica Nucleare, Sezione di Catania,  Italy}
\author{A. I. Muminov}
\address{Institute of Nuclear Physics, Tashkent, Ulugbek, 100214,
Uzbekistan}


\begin{abstract}
The mixing of the quasifission component to the fissionlike cross
section causes  ambiguity in the quantitative estimation of the
complete fusion cross section from the observed angular and mass
distributions of the binary products. We show that the
partial cross section of quasifission component of binary fragments covers
the whole range of the angular momentum values leading to capture.
The calculated angular momentum distributions for the compound nucleus and dinuclear system going to quasifission may overlap: competition between complete fusion and quasifission takes place at all values of initial orbital
angular momentum. Quasifission components formed at large angular
momentum of the dinuclear system can show isotropic angular
distribution and their mass distribution can be in mass symmetric
region similar to the characteristics of fusion-fission
components. As result the unintentional inclusion of the quasifission  contribution into  the  fusion-fission fragment yields  can lead  to overestimation of the probability of the compound nucleus formation.
\end{abstract}
\date{Today}
\maketitle

\section{Introduction}

The yield of binary events in the full momentum transfer reactions
with massive nuclei ($Z_1\times Z_2 > 1200$) at the near Coulomb barrier energies is the dominant process in comparison with the formation of the evaporation
residues. At synthesis of superheavy elements in the hot fusion
reactions with massive nuclei the authors try to choose pair of projectile-target and beam energy to reach  the maximum yield of
evaporation residues because the observed cross sections are about
or less than 1 pb (10$^{-36}$ cm$^2$) 
 \cite{OganJPhysG,OganPRC74,SHofmannEPJA32,Stavsetra,SHofmannRModPh,Morita07}.
A formation of the compound nucleus (CN) is the necessary condition
leading to yield of evaporation residues in competition with
fission. Therefore, there is a necessity to estimate as possible
the correct value of the complete fusion cross section. Because if
there is no CN we can not observe an evaporation
residue too. So, if we have the heated and rotating CN as a result of the complete fusion, the possibility to observe the evaporation residues depends on the competition between cooling by emission neutrons, protons, $\gamma$-quanta or charged particles and fission of CN. In the dinuclear system (DNS) concept \cite{Volkov} the evaporation residue cross section at collision energy  $E_{\rm c.m.}$ is factorized as follows:
\begin{equation}
\sigma_{ER}(E_{\rm c.m.})=\sum_{\ell=0}^{\ell_f}\sigma_{cap}^{(\ell)}(E_{\rm c.m.})P_{CN}^{(\ell)}(E_{\rm c.m.})
W_{sur}^{(\ell)}(E_{\rm c.m.}),
\end{equation}
where $\sigma_{cap}^{(\ell)}$ is the partial cross section of capture of
the projectile by the target nucleus; $P_{\rm CN}^{(\ell)}$ is the strength of
hindrance to formation of CN  at stage of competition between complete fusion and quasifission; $W_{\rm sur}^{(\ell)}$ is the survival probability of the heated and rotating nucleus at formation of the evaporation residue; $\ell_f$ is
the value of angular momentum at which the fission barrier of
the corresponding CN disappears: $W_{sur}^{(\ell)}(E)=0$ for $\ell > \ell_f$.
The decrease of $W_{\rm sur}$ by increasing the excitation energy
is determined by the increase of number of competition cascades
between fission and emission of particles. The synthesis of superheavy elements with $Z>113$ may be realized in the ``hot fusion" reactions at excitation energies of CN $E^*_{\rm CN}>25$ MeV which take place in collisions of  relatively light nuclei (for example, $^{48}$Ca) on transactinide targets (U, Am, Pu, Cm, Bk and Cf). In this kind of reactions the
number of formed compound nuclei are much larger than in ``cold
fusion" reactions ($E^*_{\rm CN}<25$ MeV)  where the total number of compound
nuclei is small due to the strong hindrance to complete fusion caused by quasifission. But the probability of the evaporation residue formation $W_{\rm sur}$ is large because of small values of $E^*_{\rm CN}$.
The very hard experiment to synthesis of the superheavy element $^{278}113$ in the $^{70}$Zn+$^{209}$Bi reaction have shown that the use of ``cold fusion"  reactions exhausted its potentialities \cite{Morita07}. Therefore, in the Lawrence Berkeley National Laboratory (California, USA) \cite{Stavsetra}
and GSI (Darmstadt, Germany) \cite{SHofmannCaU} the experiments
with ``hot fusion" reactions are performed  as in the Flerov
Laboratory of Nuclear Reactions (Dubna, Russia) \cite{OganJPhysG,OganPRC74}.

The study of the ``cold fusion" mechanism was relatively easy
because the target-nuclei were nearly spherical and the number of open channels is small. The number of partial waves contributing to complete fusion was not large because of small size of the potential well \cite{GiardinaEPJ8} in the interaction potential for the more symmetric system.

 The entrance channel of the  ``hot fusion" reactions is mass asymmetric.  Therefore, the size of the potential well in nucleus-nucleus interaction is larger in comparison with the one in case ``cold fusion". Consequently, the large number of partial waves can contribute to capture and complete
fusion processes. Consequently, the large lifetime of DNS at small excitation energies and the population of the large angular momentum can allow DNS to rotate on more large angles. As a result the angular distribution of quasifission products can be even isotropic. This means that if the mass distribution of quasifission products formed at decay of DNS
rotating on large angles could reach mass symmetric region, then they are confused with the products of the fusion-fission reactions. In this case we have
ambiguity in determination of the complete fusion cross section
from the measured fission products. The actinides used in ``hot fusion" reactions are deformed nuclei and the role of the orientation angle of their symmetry axis should be taken into account in study of the reaction mechanism \cite{NPA759}. The method of the calculation of quasifission contribution to the fissionlike allows us to separate pure fusion cross section from the measured cross section of the fissionlike product yields.

\section{Angular momentum distribution of fusion-fission and quasifission
products}

\begin{figure}[th]
\vspace*{-2.00 cm}
\begin{center}
\resizebox{0.85\textwidth}{!}{\includegraphics{{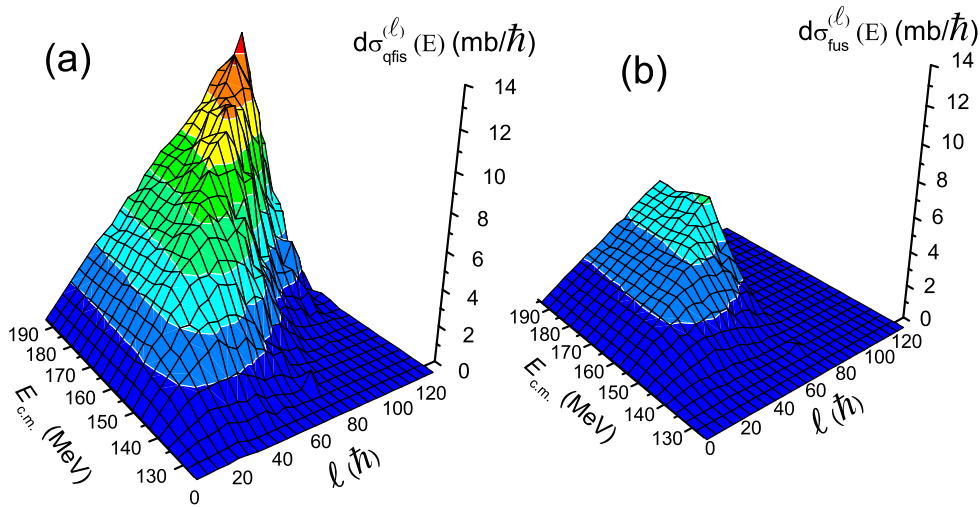}}}
\vspace*{-1.0 cm} \caption{\label{difspin} Comparison of the angular momentum
distributions  of the  quasifission (a) and fusion-fission (b)
products as a function of the initial beam energy for the
$^{48}$Ca+$^{154}$Sm reaction.}
\end{center}
\end{figure}

 Our calculations showed that the partial cross section of quasifission component of the binary fragments are distributed in the whole range of the angular momentum values leading to capture $0<\ell<\ell_{\rm max}$, where $\ell_{\rm max}$ is maximal value of $\ell$ leading to full momentum transfer (capture) reaction (see Fig. \ref{difspin}). This conclusion is different from the assumption that quasifission occurs in the range $\ell_{\rm fus}<\ell<\ell_{\rm max}$ where $\ell_{\rm fus}$ is the upper limit of angular momentum leading to complete fusion \cite{Back31}.
The calculated angular momentum distributions for the compound nucleus and dinuclear system going to quasifission may overlap: competition between complete fusion and quasifission takes place at all values of initial orbital
angular momentum.
The method based on the assumption that quasifission component is associated only with the highest partial waves can lead to incorrect quantitative results for the fusion-fission cross section and anisotropy in the angular distribution of the fusion-fission fragments.
As a result the analysis of the observed angular anisotropy of products by the transition saddle point model may be ambiguous if the contribution of the quasifission in the measured data for the fissionlike fragments is large.
This means that some part of the quasifission fragments are considered
as the fusion-fission fragments, consequently, the fusion cross
section is overestimated at analysis of the experimental data.

\section{Mixing of quasifission and fusion-fission mass distributions}

The experimental and theoretical studies showed the mass
distribution of the quasifission fragments has local maximums
around magic numbers of protons $Z$=20, 28, 50, 82 or neutrons
$N$=50, 82, 126. Total kinetic energy (TKE) distribution is very
close to the Viola systematic as for fusion-fission: TKE=$Z_1 Z_2 {\rm e}^2/D(A_1,A_2)$ or TKE may be higher if the angular momentum of
the splitting DNS was large.

\begin{figure}[th]
\vspace*{4.35 cm}
\begin{center}
\resizebox{0.80\textwidth}{!}{\includegraphics{{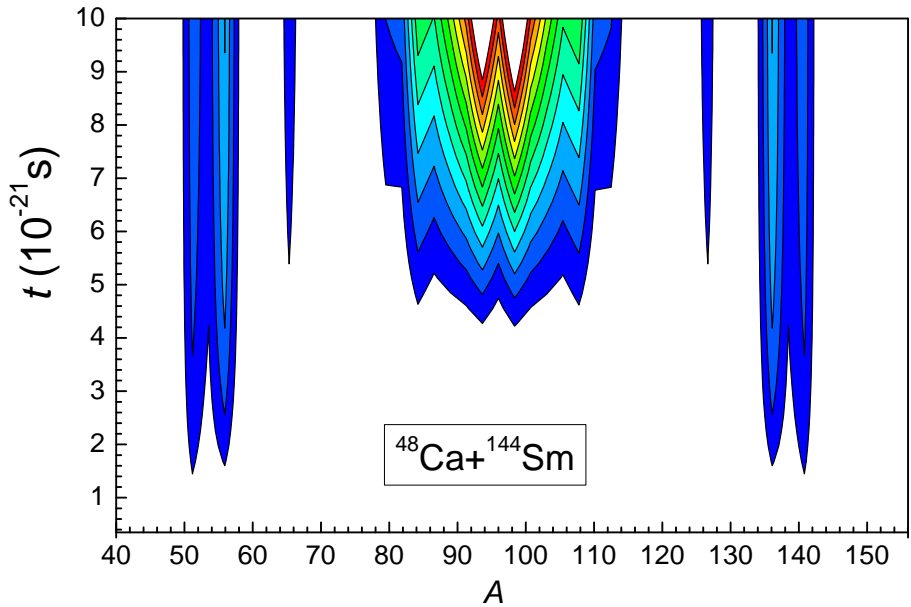}}}
\vspace*{-5.35 cm} \caption{\label{YmasdisS} Yield of the quasifission products as a function of the DNS lifetime for the $^{48}$Ca+$^{144}$Sm
reaction at $E_{\rm c.m.}=$150 MeV.} 
\end{center}
\end{figure}

\begin{figure}
\hspace*{-3.0cm}
\begin{minipage}{6.6cm}
\begin{tabular}{c}
\includegraphics[height=6.0cm]{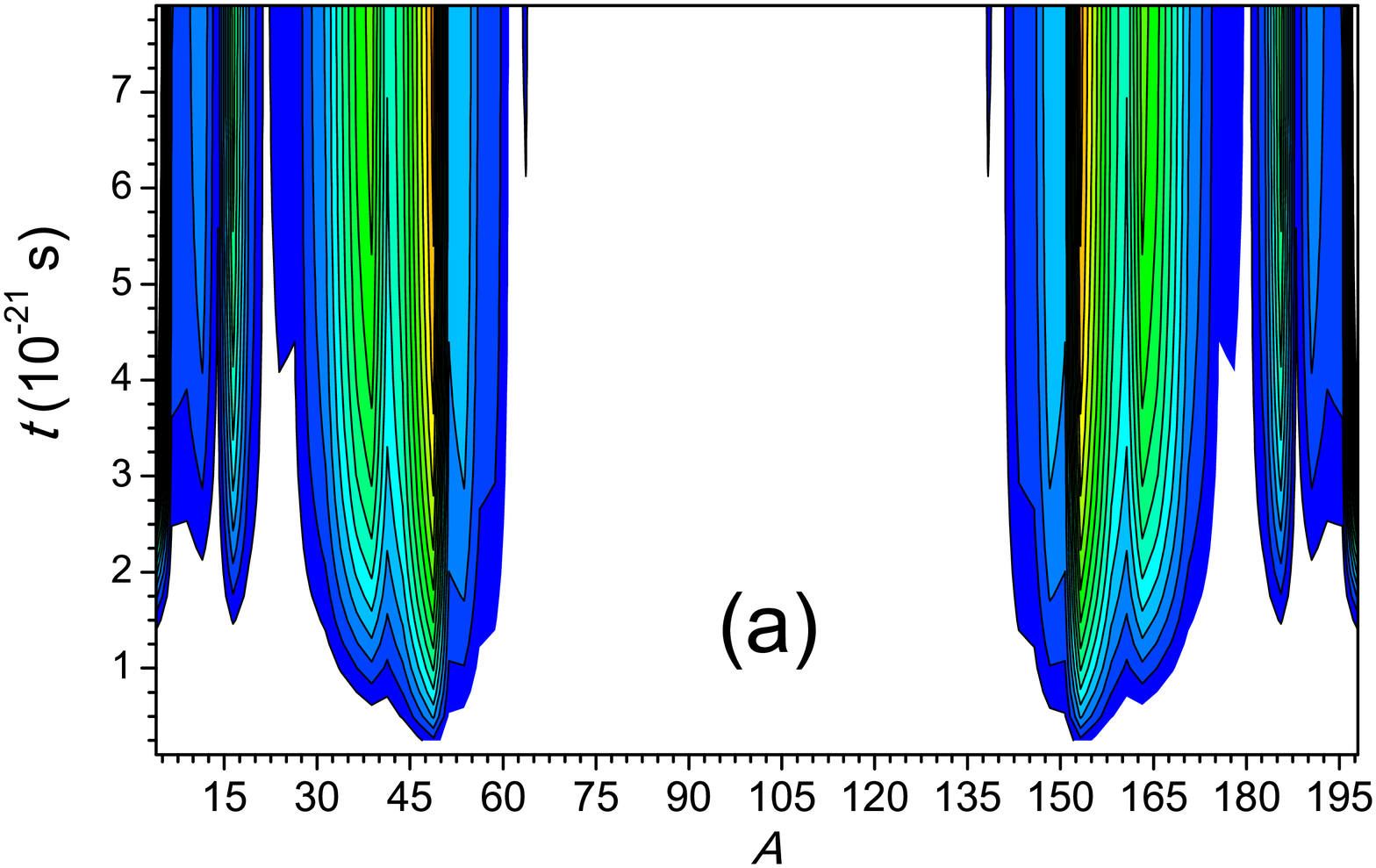}
\end{tabular}
\end{minipage}
\hspace*{1.5cm}
\begin{minipage}{6.6cm}
\begin{tabular}{c}
\includegraphics[height=6.0cm]{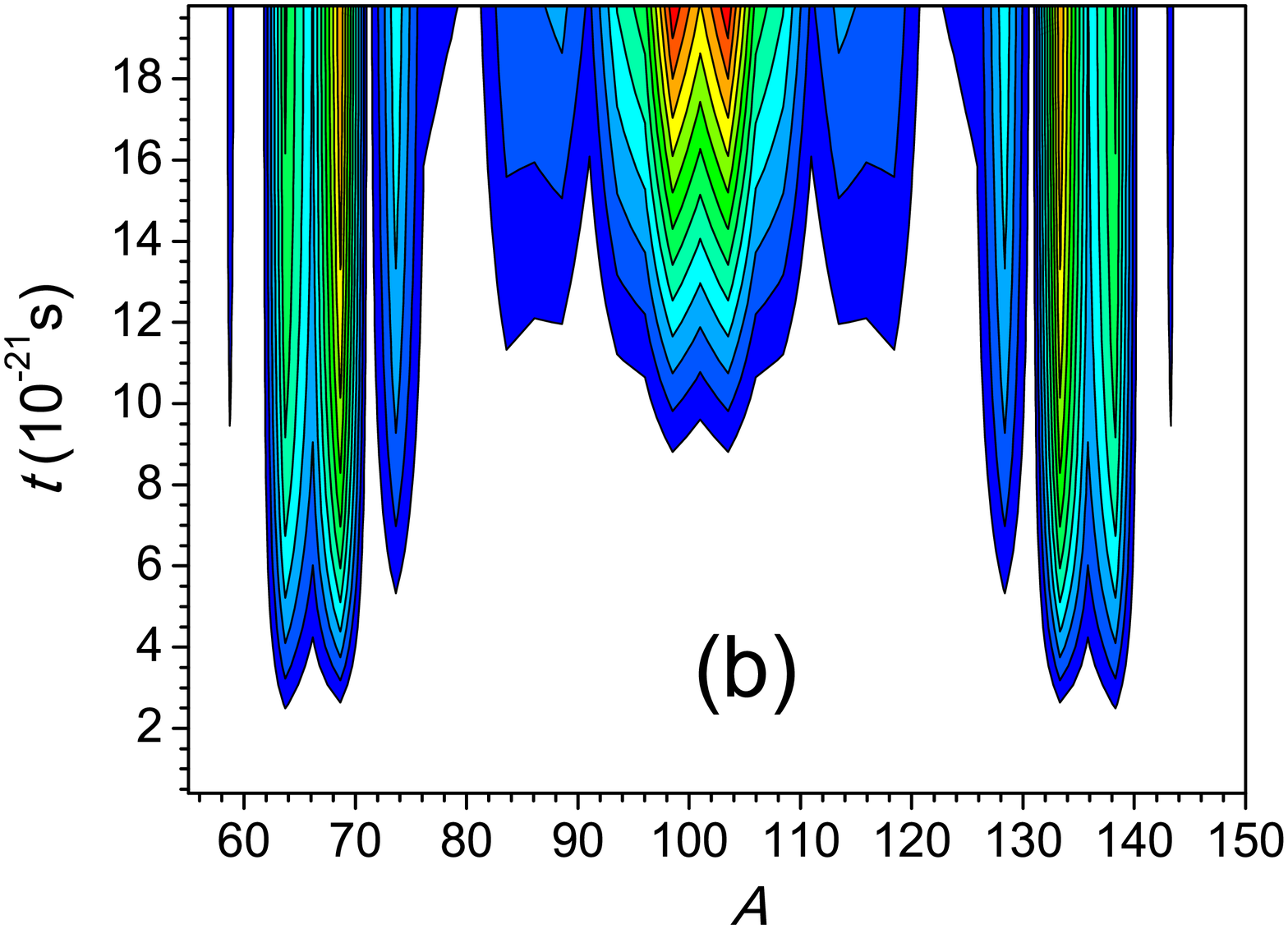}
\end{tabular}
\end{minipage}
\vspace{-0.6cm} \hspace{-1.6cm} \caption{The mass distribution of
the quasifission products yield in the $^{48}$Ca+$^{154}$Sm
reaction at $E_{\rm c.m.}$=140 MeV as a function of the lifetime
of the dinuclear system formed at capture stage (a). The mass
distribution of the quasifission product yields in the
$^{48}$Ca+$^{154}$Sm reaction at $E_{\rm c.m.}$=160 MeV as a
function of the lifetime of the dinuclear system (b).}
\label{MasDis154}
\end{figure}

Fig. \ref{YmasdisS} shows the evolution of the mass distribution as
a function of the DNS lifetime calculated in the
DNS model for the $^{48}$Ca + $^{144}$Sm reaction.  The quasifission process events are observed at reaction times longer than $1.5\relbar$2$\cdot 10^{-21}$s. The maximum of mass distribution of quasifission products can be located around or far from the initial masses of colliding nuclei in dependence on the peculiarities of the shell structure of the initial and being formed nuclei. Therefore, mass distribution of the quasifission products are different in the $^{48}$Ca + $^{144}$Sm and $^{48}$Ca + $^{154}$Sm reactions. It is seen from comparison of the corresponding experimental data presented in Ref. \cite{Knyazheva} and theoretical results obtained by us for the mass distribution evolution of the quasifission fragments which are shown in  Figs. \ref{YmasdisS} and  \ref{MasDis154}.

From the analysis of the experimental data in reactions induced by the  Ca, Ti, and Cr projectiles on the actinide targets \cite{ItkisNPA} the authors  concluded that the only peaks in the mass distribution between mass symmetric region $(A_1+A_2)/2\pm 20$ and mass of the projectile-like fragments belong to the quasifission products. As a function of the beam energy and
mass number of isotopes these peaks can change their position or
disappear completely \cite{Knyazheva}. Authors of Ref. \cite{Knyazheva} analyzed their experimental data for the quasifission  products in the $^{40,48}$Ca+$^{144,154}$Sm reactions and  for the first time they presented quantitative results about quasifission components in the yield of fissionlike
products. To clarify the role of the entrance channel characteristics in the fusion suppression and an appearance of the quasifission products the authors of Ref. \cite{Knyazheva} compared the results of the above mentioned reactions with the $^{16}$O+$^{186}$W reaction. In this reaction, a peak of fissionlike
products which is significative of the presence of quasifission
was not observed in the expected area of the mass distribution.

 The similar behaviour of the fissionlike products was observed in the $^{48}$Ca+$^{144}$Sm reaction, consequently, the authors of Ref. \cite{Knyazheva} concluded that there is no components of quasifission in this reaction too. From our point of view the quasifission components are mixed with projectile-like and target-like products.  But they excluded from the analysis the
reaction products with masses $A_1<55$ and corresponding conjugate heavy
fragments $A_{tot}-A_1$ assuming that there is no quasifission
components with masses $A_1<55$.  We think that the experimentalists underestimate capture and quasifission cross sections and, therefore, they lost  information about reaction mechanism ignoring the products with masses $A_1<55$.
There is a mixing of products of the quasifission and deep-inelastic collisions near the projectile-like and target-like fragments masses  (see Fig. 4 of Ref. \cite{faziolett}) and Figs. \ref{YmasdisS} and \ref{MasDis154} of
this paper). In Ref. \cite{faziolett} we showed that in the $^{48}$Ca + $^{208}$Pb reaction the part of the quasifission products are concentrated near masses of the projectile-like and target-like products due peculiarities of the shell structure in the colliding double magic nuclei. The part of the quasifission products mixed with the ones of deep-inelastic collisions should not be ignored at analysis of the experimental data and the attempts to extract an information from the studies of these products lead to new results about capture reactions at low energies.

The quasifission fragments can be mixed with the fusion-fission products
in the mass symmetric region.  The symmetric region is reachable for the mass distribution during evolution of DNS at more high energies before its decay into two quasifission fragments (see Fig. \ref{YmasdisS} and \ref{MasDis154}). The quasifission products having masses in the range  $(A_1+A_2)/2\pm 20$ may be
considered as the fusion-fission products. As a result of the
mixing the quasifission and fusion-fission products in the mass symmetric
region the reconstructed complete fusion cross section will be overestimated. About the possibility of mixing their angular distributions we will discuss in the next Section.

\section{Angular distribution of fissionlike fragments}
\label{angldis}

The characteristic feature for the fusion-fission products is the
isotropic angular distribution. The observed anisotropy in their
angular distribution can be described by the transition saddle
state model  \cite{Halpern,Griffin}:
\begin{equation}
\label{ALK0}
 \mathcal{A} \approx 1+ \frac{< \ell^2 >}{4 K_0^2},
 \end{equation}
where $K_0^2=\mathcal{J}_{eff}T/\hbar^2$ is the variance of the
$K$ (projection of total spin of the fissioning nucleus on its symmetry axis) distribution; $\mathcal{J}_{eff}$ is the effective moment of
inertia for the  CN and its value is determined in
the framework of the rotated finite range model by Sierk
 \cite{SierkPRC33}.
\begin{equation}
\label{Jeff}
  \frac{1}{\mathcal{J}_{eff}}= \frac{1}{\mathcal{J}_{\|}} - \frac{1}{\mathcal{J}_{\bot}}.
\end{equation}
$\mathcal{J}_{\|}$ and  $\mathcal{J}_{\bot}$ are moments of
inertia for rotations around the symmetry axis and a perpendicular
axis, respectively; the effective temperature $T$ is related to
the excitation energy $E^*$ by the expression
\begin{equation}
\label{temp}
 T=3.46 \sqrt{E^*/A}.
\end{equation}

From the other side $K_0^2$  can be extracted from the description
of angular distribution $W(\theta)$ of the fusion-fission
fragments:
\begin{equation}\label{Wtheta}
W(\theta) = \sum_{I=0}^{I_{\rm
max}}(2I+1)\frac{\sum_{L=-I}^{I}\frac{1}{2}
(2I+1)|\mathcal{D}_{M=0,K}^{I}(\theta)|^2\exp\left(-K^2/2K_0^2\right)}
{\sum_{L=-I}^{I}\exp\left(-K^2/2K_0^2\right)}
\end{equation}
which involves summations over $I$ and $K$ of the symmetric-top
wave function $\mathcal{D}_{M=0,K}^{I}(\theta)$ and assumes a
sharp-cutoff expression for the spin distribution and a Gaussian
$K$ distribution. The value of $I_{\rm max}$ used in the
analysis was determined from the measured total fission cross
sections for which the fission probability is assumed to be equal 1.
Then  the experimental value of $\mathcal{A}$ is found as the ratio of
angular distribution $W(\theta)$ value at $0^{\circ}$ to the one
at $90^{\circ}$: $\mathcal{A}=W(0^{\circ})/W(90^{\circ})$ \cite{Back}. The comparison of the experimental and theoretical
values of the  $\mathcal{J}_{0}/\mathcal{J}_{eff}$ ratio as a
function of $<I^2>$ shows that for some reactions like
$^{32}$S+$^{208}$Pb it is impossible to reach agreement between
the corresponding values \cite{BackPRL50,Back}:  the theoretical values
underestimate noticeably the experimental data ($\mathcal{J}_{0}$ is the rigid moment of inertia for a sphere of equal volume).  The authors of Ref. \cite{BackPRL50,Back} explained the deviation from the standard theoretical description of the experimental anisotropy by the contribution of quasifission
components. So it was qualitatively recognized the perceptible contribution of the quasifission  events in the yield of the $^{32}$S+$^{208}$Pb
reaction products.

The rotational angles of the DNS during capture and
before its decay into two fragments can be calculated in our model
presented in Ref. \cite{EPJA34}. For the given initial values  of
beam energy and orbital angular momentum $\ell_0$ the capture
probability is found by solving  the equation of motions \cite{NPA759,Fazio04}. If we neglect the decrease of the angular momentum
of the DNS by emission of light particles (gamma
quanta, neutrons, {\it etc.}) during its evolution to
quasifission, its angular momentum $\ell$  can be considered as a
constant value. We should note that $\ell$  is less than the
initial orbital angular momentum $\ell_0$ due to dissipation
during capture \cite{NPA759,Fazio04}. Knowing of $\ell$ and moment of
inertia $J_{(DNS)}$  of the DNS allows us to find its
angular velocity $\Omega_{DNS}$. At the considered beam energies,
the DNS is formed  when the interacting nuclei are
trapped into potential well: the relative kinetic energy decreases
due to the dissipation and it becomes not enough to overcome the
quasifission barrier by the classical dynamical way (see Fig. 2 in
Ref. \cite{NPA759}). The characteristic lifetime of DNS at
quasifission is large than $5 \cdot 10^{-22}$s which is time for the deep inelastic collisions. To find the angular distribution of the quasifission fragments we estimate the rotational angle $\theta_{DNS}$  at break-up of the system:
\begin{equation}
\theta_{DNS}=\theta_{capture}+\Omega_{DNS} \cdot
\tau(T_Z(\ell,E^*_Z(\ell))).
\end{equation}
 It can be found if we know the lifetime
($\tau(T_Z)$) of the rotating DNS which is heated up
to the effective temperature $T_Z(\ell,E^*_Z(\ell))$, where
$E^*_Z(\ell)$ is the excitation energy of DNS: $E^*_Z(\ell)=E_{\rm
c.m.}-V_{\rm min}(Z,\ell)+Q_{gg}$. Here $V_{\rm min}(Z,\ell)$ is
the minimum value of the potential pocket of the nucleus-nucleus
interaction for the given charge and mass asymmetry of DNS
fragments and $Q_{gg}=B(Z_1,A_1)+B(Z_2,A_2)-B(Z_P,A_P)-B(Z_T,A_T)$
is a change of intrinsic energy of fragments during the evolution
of DNS. $T_Z(\ell,E^*_Z(\ell))$ is calculated by formula
(\ref{temp}).
 The requested  decay time $\tau$ is estimated by
\begin{equation}
\tau(T_Z)=\frac{\hbar}{\Gamma_{qfiss}(T_Z)}
\end{equation}
if we know the excitation energy  $E^*_{DNS}$ and quasifission
barrier $B_{qf}$ of the DNS for its decay on
fragments with charge numbers $Z_1$ and $Z_2$, by using the
one-dimensional Kramers rate  \cite{Kramers,Grange,Froeb92}
\begin{eqnarray}
\Gamma_{qfiss}(\Theta)&=&\frac{K_{qf}}{K_{m}} \,\omega_m
\left(\sqrt{\gamma^2/(2\mu_{qf})^2+\omega^2_{qf}}-\gamma/(2\mu_{qf})
\right)\nonumber\\
&\times&\exp\left(-B_{qf}/T_Z)\right)/(2\pi\omega_{qf}).
\end{eqnarray}
Here the frequency $\omega_m$ and $\omega_{qf}$ are found by the
harmonic oscillator approximation to the nucleus-nucleus potential
$V(R)$ shape for the given DNS configuration $(Z_1,Z_2)$:  $\omega_m$ and
$\omega_{qf}$ describe the shapes of the potential well's bottom and
barrier placed at $R_m$ and $R_{qf}$ (see Fig. 1 of Ref.  \cite{EPJA34}),
respectively:
\begin{eqnarray}
 \omega_m^2&=&\mu_{m}^{-1}\left|\frac{\partial^2 V(R)}{\partial
R^2}\right|_{R=R_m}\,,\\
\omega_{qf}^2&=&\mu_{qf}^{-1}\left|\frac{\partial^2 V(R)}{\partial
R^2}\right|_{R=R_{qf}}.
\end{eqnarray}

\begin{figure}[th]
\begin{center}
\resizebox{0.80\textwidth}{!}{\includegraphics{{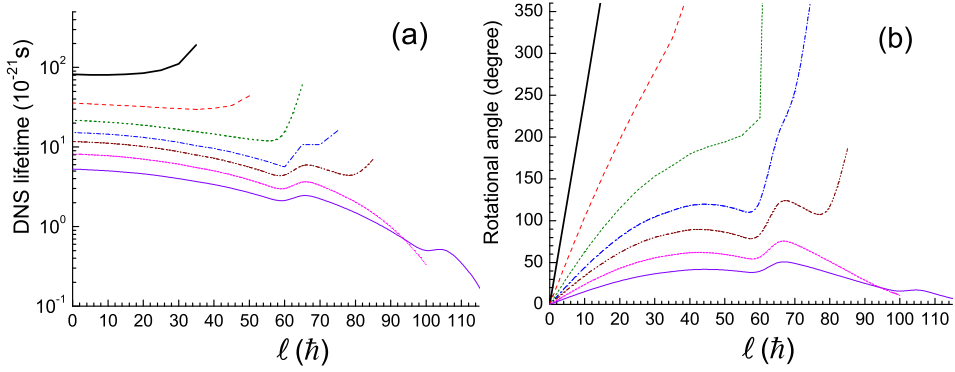}}}
\vspace*{-4.35 cm} \caption{\label{timeanglL} 
The lifetime (a) and  rotational angle
(b) of the DNS formed in  the $^{48}$Ca+$^{154}$Sm
reaction and decaying on $^{58}$Cr and $^{144}$Ce as a function of
the angular momentum for the different values of beam energies:
 $E_{\rm c.m.}$=157 (thick solid line), 160 (short dashed), 164 (dotted),
168 (dot-dashed), 171 (dot-dot-dashed), 178 (short dotted) and 191 MeV (thin solid line).}
\end{center}
\end{figure}
  We used  $\gamma=8\cdot 10^{-22}$ MeV fm$^{-2}$s in our calculations;
$\mu_{m}$ and $\mu_{qf}$ are the moment of inertia of DNS for the radial motion when internuclear distance was equal to $R_{qf}$ and $R_{m}$, respectively. Their values were found by the method presented in Ref. \cite{GiardinaEPJ8}. The corresponding collective enhancement factors  $K_{qf}$ and $K_{m}$ of
the rotational motion to the level density  are calculated by the expression suggested in
Ref.  \cite{Junghans}:
 \[K(E_{DNS}) =
 \left\{\begin{array}{ll}(\sigma_{\bot}^2-1)f(E_{DNS})+1,  \hspace*{0.2 cm}  \rm{if}
\ \  \sigma_{\bot}>1 \ \  \   \\
1, \hspace*{0.2cm} \rm{if} \ \
  \sigma_{\bot}\le 1\:,
 \end{array}
 \right.
 \]
where  $\sigma_{\bot}=J_{(DNS)} T/\hbar^2$;
$f(E)=(1+\exp[(E-E_{cr})/d_{cr}])$;  $E_{cr}=120
\widetilde{\beta}_2^2 A^{1/3}$ MeV; $d_{cr}=1400
\widetilde{\beta}_2^2 A^{2/3}$. $\widetilde{\beta}$ is the
effective quadrupole deformation for the DNS. Its value was calculated from the estimation of  $\mathcal{J}^{(DNS)}_{\bot}$  for the shape of DNS corresponding to the minimum and  maximum of potential energy taken as a function of the relative distance $R$  between centers of fragments of DNS. $J_{(DNS)}$ is the moment of inertia of DNS around the axis which is  perpendicular to $R$.

 The results of calculations of the DNS lifetime and
 angular distributions of the quasifission fragments $^{58}$Cr and
 $^{144}$Ce for the different values of
 the beam energy and angular momentum of DNS are shown in
 Figs. \ref{timeanglL}a and \ref{timeanglL}b, respectively.
 We can see that the DNS lifetime increases allowing fragments to be
 distributed to large angles  by decreasing of the beam energy.
 As expected the lifetime of DNS and angular distribution of quasifission
 fragments depend on the values of $\ell$ too. The yields of fragments with different masses  (charges) and their angular distributions are in the strong dependence on the parameters of decay channel as quasifission barrier $B_{\rm qf}$ and excitation energy of DNS with the given mass and charge asymmetry.
The analysis of this variety of dependencies of the mass and angular distribution of the quasifission products allows us to explain the reasons causing  the authors of Ref.  \cite{Knyazheva} to conclude about decrease
 of quasifission phenomenon by increase of beam energy in the
 $^{48}$Ca+$^{154}$Sm reaction.

\section{Dependence of mass and angular distributions
 on the beam energy}

\begin{figure}[th]
\begin{center}
\resizebox{0.75\textwidth}{!}{\includegraphics{{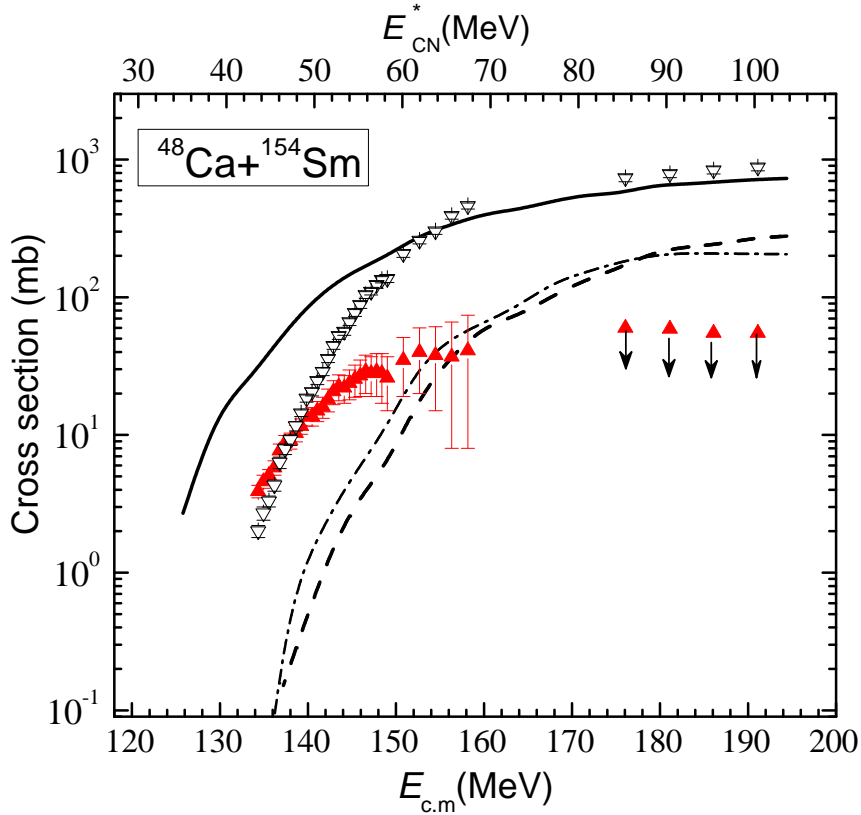}}}
\vspace*{-7.15 cm} \caption{\label{compcross} 
Comparison of the theoretical results
obtained in the DNS model (curves) \protect \cite{Fazio04,NPA759}
and experimental (symbols) \protect \cite{Knyazheva} excitation
functions for the quasifission (solid line and solid triangles),fusion-fission (dot-dashed line and inverted open triangles)  and
fast fission (dashed line) for the $^{48}$Ca+$^{154}$Sm reaction.
The theoretical values of the fusion-fission cross section are
calculated by the advanced statistical model
\protect \cite{DArrigo} } 
\end{center}
\end{figure}

From the experimental studies
 \cite{Knyazheva,ItkisNPA,Toke85,Shen87} is known that the ratio
between the quasifission and fusion-fission components in the
fissionlike products, the positions of the maximum of the mass
and angular distributions of the quasifission fragments for
given reaction depend on the initial collision energy $E_{\rm c.m.}$.
This is seen from Fig.\ref{compcross} where we compare  the theoretical results obtained in the DNS model (curves)  \cite{Fazio04,NPA759} and experimental
(symbols)  \cite{Knyazheva} excitation functions for the
quasifission (solid line and solid triangles), fusion-fission
(dot-dashed line and inverted open triangles)  and fast fission
(dashed line) for the $^{48}$Ca+$^{154}$Sm reaction. The theoretical results
(solid line) overestimate noticeably the experimental quasifission cross section (solid triangles). How this difference can be explained?

The first reason is the exclusion of the fissionlike products having
mass numbers outside of the mass range $55<A<145$  from their
analysis by the authors of Ref.  \cite{Knyazheva}.

 Consequently they lost a part of the capture cross
sections $\sigma^{(\rm exp)}_{\rm cap}$ related to the
contributions of the  quasifission fragments $\sigma^{(\rm
exp)}_{\rm qf}$ with $A_{\rm qf}<55$ and $A_{\rm qf}>145$ at the given
collision energy $E=E_{\rm c.m.}$:
\begin{equation}
   \sigma^{(\rm exp)}_{\rm cap}(E,A_{\rm qf})=
   \sigma^{(\rm exp)}_{\rm ER}(E)+
   \sigma^{(\rm exp)}_{\rm f}(E)+
   \sigma^{(\rm exp)}_{\rm qf}(E, 55<A_{\rm qf}<145),
\end{equation}
where $\sigma^{(\rm exp)}_{\rm f}$ is the cross section of the
measured yield of fission products. We described the
excitation functions of evaporation residues obtained from
Ref.  \cite{Stefanini} by our model using the
theoretical cross section of capture events which included the
contributions of all fragment yields, {\it i.e.} $4<A_{\rm
qf}<198$, from full momentum transfer reactions:
\begin{equation}
   \sigma_{\rm cap}(E)=
   \sigma_{\rm ER}(E)+
   \sigma_{\rm f}(E)+
   \sigma_{\rm qf}(E)+
   \sigma_{\rm fast-fission}(E).
\end{equation}

Because our studies showed that the full momentum transfer events
lead to the yield of fragments with masses $A_{\rm qf}<55$ too.
It is seen in the left panel of Fig. \ref{MasDis154} where we
present the evolution of mass distribution of the quasifission
products  which were calculated by our
method presented in Ref. \cite{faziolett}. The observed
quasifission feature at low energies is connected with the
peculiarities of the shell structure of the interacting nuclei.

The experimental and theoretical quasifission cross sections do
not come closer by increase in the beam energy in spite of the amount of
the quasifission fragments with masses in the $70<A_{\rm qf}<130$ range
into the measured fission events increases. But the increase
of beam energy leads to formation of DNS with large angular
momentum and the products of its decay can have isotropic
angular distribution. Therefore, such events were considered
as a fusion-fission fragments although they belong to quasifission.
Some part of the DNS mass distribution moves to the mass symmetric region due to the increase of excitation energy of DNS. Our theoretical results showing the increase in the quasifission fragment
yields in the mass symmetric region are presented in the right panel
of Fig. \ref{MasDis154}.

As we can see from Fig. \ref{compcross} that the sum of the experimental values of the fusion-fission and quasifission cross sections are very close
to our theoretical results at $E_{\rm c.m.} > 150 $ MeV although the proportion of the contributions are different. The mixing of the quasifission and
fusion-fission products causes ambiguity at the reconstruction of the
complete fusion cross section from the registered fission products. This is a reason why our theoretical results for quasifission events overestimate the corresponding experimental data and why our theoretical results for
fusion-fission products underestimate the corresponding data
presented by authors.

\section*{Conclusions}

We showed that the partial cross section of quasifission component of binary fragments are distributed in the whole range of the angular momentum values leading to capture $0<\ell<\ell_{\rm max}$.  This means that the angular momentum distributions of the compound nucleus and dinuclear system going to quasifission  overlap because competition between complete fusion and quasifission takes place at all values of initial orbital angular momentum. Our conclusion is different from the assumption that quasifission occurs in the range $\ell_{\rm fus}<\ell<\ell_{\rm max}$ where $\ell_{\rm fus}$ is the upper limit of angular momentum leading to complete fusion  \cite{Back31}. But the partial cross section of fast fission process is distributed in the angular momentum range $\ell_{\rm f}<\ell<\ell_{\rm max}$ where the fission barrier for the CN is equal to zero.
Calculations of lifetime of the rotating dinuclear system demonstrated that at the near Coulomb barrier energies the angular distribution of the quasifission fragments with the projectile-like masses can reach large angles. The maximum of the angular distribution moves to the forward (backward for conjugate fragment) angles by the increase of the beam energy due to the decrease of the lifetime of the dinuclear system. The calculation of the mass distribution of the quasifission products showed that during its evolution it can reach the mass symmetric region. So, if the angular distribution fragments of this region becomes isotropic then it is impossible to separate them from fusion-fission components. As result the quasifission components which could be considered as fusion-fission products lead to overestimation of the probability of the compound nucleus formation.

\section*{Acknowledgements}

This work was performed partially under the financial support of
the RFBR, INFN-BLTP agreement and  Bogoliubov-Infeld programm
which are thanked very much by the author AKN. Authors AKN and GG
are also grateful to the Fondazione Bonino-Pulejo of Messina for
the support received in the collaboration between the Dubna and
Messina groups.

\end{document}